\documentclass[11pt]{article} 
\newcommand{\be}{\begin{equation}}
\newcommand{\ee}{\end{equation}}
\newcommand{\bea}{\begin{eqnarray}}
\newcommand{\eea}{\end{eqnarray}}
\newcommand{\sn}{{\rm sn}}
\newcommand{\ds}{{\rm ds}}
\newcommand{\cs}{{\rm cs}}
\newcommand{\ns}{{\rm ns}}
\newcommand{\dn}{{\rm dn}}
\newcommand{\cn}{{\rm cn}}
\newcommand{\sech}{{\rm sech}}

\begin{document}
\vspace{.5in} 
\begin{center} 
{\LARGE{\bf Reply to the Comment by Zhang and Li [J. Math. Phys. 56, 084101, 
2015] on the Paper [J. Math. Phys. {\bf 56}, 032104, 2015]}}
\end{center} 

\vspace{.3in}
\begin{center} 
{\LARGE{\bf Avinash Khare}} \\ 
{INSA Senior Scientist,}
{Physics Department, Savitribai Phule Pune University, \\
Pune, India 411007}
\end{center} 

\begin{center} 
{\LARGE{\bf Avadh Saxena}} \\ 
{Theoretical Division and Center for Nonlinear Studies, Los
Alamos National Laboratory, Los Alamos, NM 87545, USA}
\end{center} 

\vspace{.9in}
{\bf {Abstract:}}  

In a recent paper Zhang and Li have doubted our claim that whenever a nonlinear 
equation has solutions in terms of the Jacobi elliptic functions $\cn(x,m)$ and $\dn(x,m)$, 
then the same nonlinear equation will necessarily also have solutions in terms of
$\dn(x,m) \pm \sqrt{m} \cn(x,m)$. We point out the flaw in their argument and
show why our assertion is indeed valid.

\newpage 
  
 
In a series of recent papers we had shown through a large number of examples 
\cite{ks1,ks2,ks3} that whenever a
nonlinear equation, discrete or continuous, integrable or nonintegrable, 
coupled or uncoupled, local or nonlocal has periodic solutions in terms
of Jacobi elliptic functions $\dn(x,m)$ and $\cn(x,m)$, then the same
equation will necessarily have solutions in terms of their linear
combinations $\dn(x,m) \pm \sqrt{m} \cn(x,m)$. Recently Zhang and Li 
\cite{zl} have examined our claim in the context of nonlinear Schr\"odinger
equation and have claimed that such an assertion is impossible. The 
purpose of this reply is to point out a serious flaw in their argument.

Zhang and Li start with the nonlinear Schr\"odinger equation (NLSE)
\be\label{1}
iu_t +u_{xx} + b |u|^2 u =0\,.
\ee
On assuming 
\be\label{2}
u(x,t) = e^{i(kx-\omega t) \phi(\zeta)}\,,~~\zeta = x-vt\,,
\ee  
they reduce the  NLSE to
\be\label{3}
\phi''(\zeta) = a\phi +b \phi^3\,,
\ee
where $a = \omega -k^2\,, v=2k$. Then they went on to obtain various
solutions to the Eq. (\ref{3}) including those in terms of $\dn(x,m)$ and 
$\cn(x,m)$.  Up to this point we completely agree with Zhang and Li 
\cite{zl}. 

The crucial flaw in their argument came at the stage when they asserted 
that if the nonlinear Eq. (\ref{3}) has two solutions
$\phi_1(\zeta)$ and $\phi_2(\zeta)$ then $\Phi(\zeta) = \phi_1(\zeta)
+\phi_2(\zeta)$ can be a solution of the nonlinear Eq. (\ref{3}) only if
\be\label{4}
\phi_{1}^{2} \phi_2 + \phi_{2}^{2} \phi_1 = 0\,.
\ee
This assertion is incorrect. In particular, if $\phi_1$ and $\phi_2$ are
such that 
\be\label{5}
\phi_{2}^{2} =  \phi_{1}^{2} + c\,,
\ee
then also $\phi(\zeta)$ can be a solution of Eq. (\ref{3}), where $c$ is a 
constant. And this is precisely true in the case of the elliptic functions 
$\dn(\zeta,m)$ and $\sqrt{m}\cn(\zeta,m)$. In particular, while $\dn(\zeta,m)$ 
and $\cn(\zeta,m)$ are distinct periodic functions with periods $2K(m)$ and 
$4K(m)$ (where $K(m)$ is complete elliptic integral of the first kind \cite{as}), 
they satisfy the identity
\be\label{6}
\dn^2(x,m) = m \cn^2(x,m)+ (1-m)\,.
\ee  
And precisely because of such an identity that our assertion, as has been
proved by numerous examples, is indeed correct.

One of us (AK) is grateful to Indian National Science Academy (INSA) for 
the award of INSA senior Scientist position at Savitribai Phule Pune University. 
This work was supported in part by the U.S. Department of Energy.


\begin{thebibliography}{99}

\bibitem{ks1} A. Khare and A. Saxena, J. Math. Phys. {\bf 56} (2015). 032104. 

\bibitem{ks2} A. Khare and A. Saxena, J. Math. Phys. {\bf 55} (2014) 032701. 

\bibitem{ks3} A. Khare and A. Saxena,  Phys. Lett. {\bf A377} (2013) 2761.

\bibitem{zl} Y. Zhang and J-B. Li, J. Math. Phys. {\bf 56} (2015) 084101.

\bibitem{as} M. Abramowitz and I. A. Stegun, Handbook of Mathematical
Functions (Dover, New York 2010).

\end{thebibliography}
\end{document}